# MetacouplingLLM: Integrating geographic adjacency, flow indicators, and LLM-assisted analysis

Xiang Yu[1,2,*], Yingjie Li[3], Zihan Zheng[4], Nan Jia[5], Xin Lan[1], Jianguo Liu[1]

[1] Center for Systems Integration and Sustainability, Michigan State University, East Lansing, MI, USA

[2] Environmental Science and Policy Program, Michigan State University, East Lansing, MI 48823, USA.

[3] Natural Capital Project, Woods Institute for the Environment, Doerr School of Sustainability, Stanford University, Stanford, CA, USA.

[4] Department of Civil and Environmental Engineering, University of Illinois Urbana-Champaign, Urbana, IL, USA.

[5] College of Public Administration, Nanjing Agricultural University, Nanjing 210095, China.

[*] Correspondence: yuxiang4@msu.edu

## Abstract

The metacoupling framework examines interactions within coupled human and natural systems and between adjacent or distant systems, but its application is constrained by inconsistent geographic classifications and limited standardized quantitative workflows. We present MetacouplingLLM, an open-source Python package integrating adjacency databases, reproducible flow indicators, and large language model (LLM)-assisted qualitative analysis. The databases combine administrative boundaries, hydrography, and fixed-crossing evidence, with configurable adjacency policies and optional country-level maritime adjacency. Metacoupled Flow Shares, Metacoupled Flow Evenness, and the Metacoupled Flow Concentration Index characterize flow composition, balance across coupling types, and partner concentration within each type. The qualitative workflow combines literature retrieval, optional web evidence, and geographic checks to draft structured analyses. Applications to global marine fishing and Indonesian palm oil demonstrate these capabilities; the fisheries reanalysis reveals differing partner concentration among regions with similar telecoupled flow shares. MetacouplingLLM facilitates consistent framework application and comparative sustainability research.

## 1. Introduction

Research on coupled human and natural systems emerged from the recognition that social and ecological processes are reciprocally linked and are best studied as integrated systems rather than as separate domains (Liu et al., 2007). The telecoupling framework formalized socioeconomic and environmental interactions between geographically distant systems, organizing them around five components: systems, flows, agents, causes, and effects (Liu et al., 2013; Liu et al., 2015). The metacoupling framework generalized this perspective into a single schema spanning three coupling types relative to a focal system: intracoupling within a system, pericoupling between adjacent systems, and telecoupling between distant systems (Liu, 2017). Because it captures human–nature interactions within systems and between adjacent or distant systems, the metacoupling framework has been applied broadly across sustainability science (Hull and Liu, 2018; Hovis et al., 2021; Liu, 2023).

The telecoupling and metacoupling frameworks have been applied across diverse places, sectors, and issues, including Pacific tuna ecosystem services (Drakou et al., 2018), soybean production in Brazil (Bicudo da Silva et al., 2021), global virtual water trade (Du et al., 2022), and nature-based tourism in protected areas (Chung et al., 2018). This breadth is a major strength, but it also creates practical difficulties. Each application must define the systems, flows, agents, causes, and effects of its own case, which is challenging for researchers who are unfamiliar with the framework. A review of the first five years of telecoupling research found that the framework's specific language was adopted unevenly, with 37% of the reviewed studies not using it (Kapsar et al., 2019). Quantitative applications have drawn on agent-based models (Dou et al., 2020a), network and input-output methods (Prell et al., 2017; Schaffer-Smith et al., 2018), and integrated economic-land models (Millington et al., 2017). These approaches, however, are typically developed for specific applications, leaving the metacoupling framework lacking unified quantitative methods. These gaps hinder standardization and make metacoupling studies difficult to compare and reproduce.

Determining adjacency is another recurring challenge in identifying pericoupling. Coupling type is not an intrinsic property of a pair of systems: it depends on whether the systems are adjacent or distant relative to a chosen focal system and spatial scale. Demand from distant systems can reshape interactions between adjacent systems (Herzberger et al., 2019). Spillover systems, which mediate or absorb flows between sending and receiving systems, may be geographically adjacent or distant, with potentially different consequences (Liu et al., 2018; Dou et al., 2020b). Furthermore, because trade, migration, and investment often couple distant places more tightly than neighboring ones, the intuition behind Tobler's first law of geography does not hold for many coupling phenomena (Tobler, 1970; Manning et al., 2023). In practice, adjacency is also ambiguous where borders follow inland water or seas, where territory is disputed, and at subnational scales, and no single boundary or hydrographic dataset resolves these cases. Determining whether two systems are adjacent is therefore a substantive analytical step, which requires an explicit and auditable geographic reference.

Existing software supports only parts of the metacoupling workflow and generally assumes that the conceptual framing has already been established. The Telecoupling Toolbox and the Telecoupling GeoApp provide GIS-based environments for mapping and quantifying the five framework components from user-supplied spatial layers (Tonini and Liu, 2017; McCord et al., 2018). However, both focus primarily on telecoupling analyses, which require users to define the relevant systems in advance, and do not support natural-language research descriptions as inputs. Large language models (LLMs) have demonstrated capabilities in extracting and synthesizing information from scientific text (Dagdelen et al., 2024; Babaei Giglou et al., 2024). However, general information extraction does not inherently incorporate framework-specific concepts or geographic validation, while unvalidated LLM outputs may contain fabricated quantitative claims or geographic relationships (Ji et al., 2023).

To address these gaps, we developed MetacouplingLLM, an open-source Python package that operationalizes the metacoupling framework through two complementary analytical workflows. The first provides an LLM-assisted qualitative workflow that turns a plain-language research topic into a structured metacoupling analysis with citations, maps, and an execution trace. This workflow draws on a corpus of 421 telecoupling and metacoupling studies and uses built-in adjacency databases to check spatial relationships, supporting structured preliminary analyses. The second provides reproducible quantitative analysis through geographic classification and flow-indicator calculation. From user-supplied flow tables, the package can compute three indicators introduced in this study: Metacoupled Flow Shares (MFS), Metacoupled Flow Evenness (MFE), and the Metacoupled Flow Concentration Index (MFCI). The geographic databases included in the package cover national- and subnational-level land and inland-water borders, with an optional country-level maritime layer. The databases support both geographic checks in the qualitative workflow and coupling classification for quantitative analysis. Together, these workflows combine reproducible geographic and quantitative methods with auditable LLM-assisted analysis that incorporates evidence retrieval, structured parsing, and adjacency checks.

The remainder of this paper is organized as follows. Section 2 describes the software architecture and safeguards for LLM-assisted analysis. Section 3 presents the national and subnational pericoupling databases, including their construction and validation. Section 4 introduces the LLM-assisted workflow and demonstrates its application to palm oil production and exports in Indonesia. Section 5 defines the metacoupled flow indicators and applies them to global marine fishing. Section 6 discusses the contributions and limitations of the package, and Section 7 presents the conclusions.

## 2. Software design and characteristics

MetacouplingLLM is an open-source Python package distributed under the MIT license, with source code and documentation available on GitHub (https://github.com/Hanshan-XY/MetacouplingLLM). MetacouplingLLM version 0.3.0 requires Python 3.10 or later. The `MetacouplingAssistant` class serves as the primary interface for qualitative analysis, whereas quantitative indicators are implemented in the `metacouplingllm.indicators` module. Core dependencies include NumPy and FastEmbed, with optional dependencies supporting LLM integration, visualization, web search, Word document export, and flow-indicator calculation. Fig. 1 illustrates the package architecture.

MetacouplingLLM supports two complementary research workflows (Fig. 1a). The first supports researchers, particularly those new to the metacoupling framework, by providing structured preliminary analyses, literature recommendations, evidence from web sources, maps, and exportable reports. The second enables researchers with existing flow data to calculate reproducible indicators of flow distribution across coupling types and concentration among partners. To support these workflows, the package separates reproducible computation from LLM-generated content (Section 2.1) and incorporates safeguards for LLM-assisted analysis (Section 2.2).

### *2.1 Separation of reproducible and LLM-generated components*

The core design principle is that numerical results and adjacency classifications remain independent of LLMs unless users explicitly request model-assisted labeling. In the default quantitative workflow, users supply flow magnitudes and either provide coupling-type labels or use the package to derive them from the specified focal system and adjacency databases. The package then computes the indicators using pandas and NumPy, ensuring that identical input tables produce identical indicator values.

LLMs are used for tasks involving language processing and judgment. In the qualitative workflow, `MetacouplingAssistant` uses an LLM to draft structured analyses (Section 4). For flow-indicator calculation, the optional `metacouplingllm.indicators.llm` submodule provides five helper functions for defining the study scope, checking inputs, classifying ambiguous flows, interpreting results, and drafting methods text (Section 5.3). These functions do not compute indicator values. Even though the `classify_ambiguous_edges` helper uses an LLM to label unresolved flows, indicator calculations remain reproducible for a fixed labeled input table.

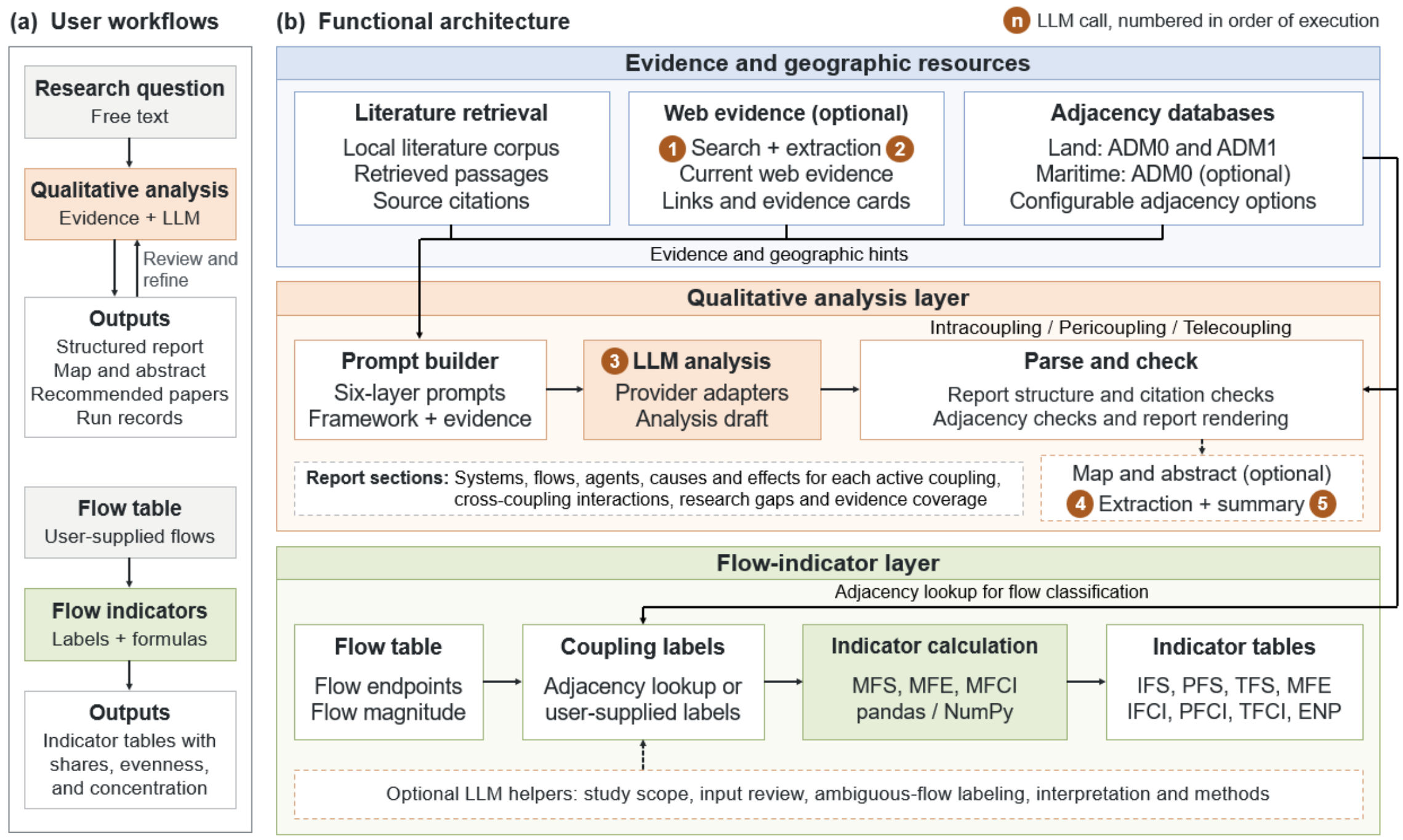

**Fig. 1. User workflows and functional architecture of MetacouplingLLM.** (a) User workflows for LLM-assisted qualitative analysis and reproducible flow-indicator calculation. (b) Functional architecture linking evidence sources, adjacency databases, qualitative analysis, and indicator computation. Numbered circles indicate the execution order of LLM-assisted steps when enabled. Orange dashed outlines denote optional LLM tasks and helpers.

### *2.2 Safeguards for LLM use*

LLMs play a central role in the qualitative workflow. Depending on the configuration, each analysis involves up to five LLM calls for web search, structured extraction of web evidence, main framework analysis, map data extraction, and abstract generation (Fig. 1b). However, LLMs may produce plausible but factually incorrect claims (Ji et al., 2023; Kalai and Vempala, 2024), express unwarranted confidence in erroneous outputs (Geng et al., 2024), and exhibit limitations in geographic reasoning (Xu et al., 2025). Their output may also be sensitive to minor changes in prompt formatting (Sclar et al., 2024). To address these limitations, the package implements three safeguards. First, the main analysis prompt incorporates supporting evidence, including adjacency information from the national or subnational databases for systems identified in the research question, retrieved literature passages, and web-search results when enabled. This encourages the model to draw on supplied evidence rather than relying exclusively on its pretrained knowledge. Second, after the main analysis is parsed, the package checks the model-generated coupling classifications against the reproducible adjacency databases and flags discrepancies for manual review. Third, the built-in execution trace records prompts and responses for the main analysis, map-data extraction, and abstract generation, together with web-search results and structured web-extraction outputs. It also records configuration parameters and the package version. These records support the auditability of individual analyses even when model outputs vary across runs (Siddiq et al., 2025). Together, these safeguards establish an LLM-assisted workflow in which evidence is supplied to the model, coupling classifications are checked against reproducible data, and execution details are recorded for subsequent review.

### *2.3 Model providers, web search, and literature retrieval*

MetacouplingLLM supports OpenAI, Anthropic, Gemini, Grok, and custom clients through a unified adapter interface. When web search is enabled (`web_search=True`), each built-in adapter uses its provider's native web-search capabilities. For custom clients, or when native search fails, the package falls back to DuckDuckGo without requiring additional dependencies. Web search provides access to current, case-specific information beyond the static literature corpus.

For literature grounding, MetacouplingLLM incorporates a corpus of 421 telecoupling and metacoupling studies from 2013 to 2025. The corpus includes the full text of 219 papers and structured paraphrases of the remaining 202 papers, for which full-text redistribution is restricted. By default, retrieval uses BAAI/bge-base-en-v1.5 embeddings through FastEmbed, with precomputed corpus embeddings bundled with the package. If the embedding backend cannot be initialized, or if lexical retrieval is explicitly selected, the package uses term frequency–inverse document frequency (TF-IDF) scoring. Retrieved passages are incorporated into the analysis prompt, and the model is instructed to cite supporting passages using turn-specific citation markers to distinguish references across multi-turn conversations. This approach follows the principles of retrieval-augmented generation, which supplies external evidence to support knowledge-intensive tasks (Lewis et al., 2020; Gao et al., 2023). It is designed to improve transparency and reliability while reducing reliance on the model's pretrained knowledge.

## 3. Pericoupling database

Pericoupling refers to interactions between geographically adjacent systems (Liu, 2017). Because adjacency is scale-dependent, adjacency at the national level does not necessarily imply adjacency between individual subnational units. Shared rivers and lakes introduce additional complexity, as water-only borders may lack fixed crossings that facilitate movement and connectivity between neighboring systems. Countries without a shared land border may nevertheless be adjacent at sea, making maritime boundaries relevant to the classification of marine interactions. Disputed territories further complicate adjacency classification because de facto territorial control may differ from legal sovereignty. To address these challenges, MetacouplingLLM uses built-in national and subnational adjacency databases to assess geographic relationships. Database lookups are reproducible under the same data version and configuration, with three configurable options (`de_facto_borders`, `coupling_standard`, and `maritime_adjacency`) controlling how disputed boundaries, water-only adjacencies, and maritime boundaries are treated.

### *3.1 Source, contiguity rule, and provenance chain*

The package provides national (ADM0) and subnational (ADM1) pericoupling databases derived from the World Bank Official Boundaries dataset (World Bank, 2026). Because this dataset does not explicitly identify inland water features, four hydrographic layers are used to identify candidate borders that follow rivers or lakes: Natural Earth river centerlines and lake polygons (Natural Earth, 2022), HydroRIVERS (Lehner and Grill, 2013), and HydroLAKES (Messager et al., 2016). HydroRIVERS and HydroLAKES supplement Natural Earth by providing coverage of smaller rivers and lakes. These hydrographic layers identify candidate water-only borders but do not independently determine their classification. Final classifications are established through case-by-case adjudication (Section 3.3). OpenStreetMap data are used to assess fixed-crossing status along water-only borders. A maritime layer incorporates data from Marine Regions.

Administrative polygons are first clipped to land using the World Bank ocean mask to minimize shoreline inconsistencies between datasets. Adjacency is then defined using rook contiguity: two administrative units are considered adjacent if they share a boundary segment of non-zero length, whereas contact at a single vertex does not qualify (Anselin et al., 2006; Rey and Anselin, 2007). A spatial index identifies candidate neighboring pairs, after which their shared boundaries are computed through geometric intersection and their lengths measured geodesically.

The ADM1 adjacency edge list is constructed in four stages (Table 1). Stage 1 derives adjacency from boundary geometry after two types of reviewed source correction: reassigning misattributed sliver corridors along international borders and reuniting an administrative unit split in the source data. Stage 2 incorporates de facto connections across disputed territories (Section 3.2). Stage 3 classifies water-only borders and adds water-separated border pairs whose polygons do not directly touch (Section 3.3). Stage 4 applies two reviewed correction files to restore five genuine land-border pairs and remove seven false contacts. The resulting ADM1 database contains 8,462 border pairs connecting 3,374 regions across 196 countries and territories. The ADM0 adjacency matrix is constructed directly from country polygons rather than aggregated from ADM1. It contains 323 exact-contact pairs and three de facto pairs, yielding a total of 326.

**Table 1. Main stages in constructing the ADM1 adjacency database.**

| Stage | Purpose | Main operations |
|---|---|---|
| 1. Boundary geometry | Establish geometry-based adjacency | Apply reviewed source corrections, identify shared boundary segments, and measure their lengths |
| 2. Disputed territories | Represent optional de facto connectivity | Derive additional connections from disputed-area polygons and their administrators |
| 3. Water classification | Identify water-only borders and fixed-crossing status | Screen borders against hydrographic layers, review candidates and crossings, and confirm water-border pairs between non-touching polygons |
| 4. Edge corrections | Restore missing borders and remove false contacts | Add 5 reviewed land-border pairs and remove 7 reviewed false contacts |

### *3.2 Disputed-area, water-only, and maritime border toggles*

The standard World Bank boundary layers exclude 24 disputed areas, creating gaps between administrative units connected through these territories. The `de_facto_borders` option determines whether the corresponding adjacencies are included and is enabled by default (`de_facto_borders=True`). When enabled, the overlay assigns each disputed area to the administrator identified for its largest portion in Natural Earth, thereby representing de facto connectivity. When disabled, the database retains the standard World Bank boundaries. This option affects 3 ADM0 pairs and 16 ADM1 pairs. As the World Bank disputed-areas layer does not identify administering countries, the overlay uses country-level administrative assignments derived from Natural Earth. With this option enabled, the ADM0 and ADM1 databases contain 326 and 8,462 border pairs, respectively.

The second option, `coupling_standard`, determines how inland water-only borders between administrative units are treated. The default lenient setting (`coupling_standard='lenient'`) retains all water-only pairs, whereas the stringent setting (`coupling_standard='stringent'`) excludes them. The moderate setting (`coupling_standard='moderate'`) retains a water-only pair only when the units are connected by at least one completed fixed crossing, including a road or railway bridge, causeway, dam-top road, or tunnel. Ferry services do not meet this criterion. This approach uses fixed transport infrastructure as an indicator of physical connectivity. For example, Kinshasa and Brazzaville are separated by the Congo River and connected only by ferry services. Therefore, their adjacency is excluded under the moderate setting. The same criterion applies to lake boundaries. For example, Flevoland and Noord-Holland are retained under the moderate setting because they are connected by the Hollandse Brug. An ADM0 pair is classified as water-only when all constituent ADM1 borders between the two countries are water-only. For such pairs, a fixed crossing is recorded if at least one constituent ADM1 border has a qualifying crossing. Table 2 summarizes the border pairs retained under each setting, while Section 3.3 describes the procedures used to identify water-only borders and qualifying fixed crossings.

The third option enables national-level maritime adjacency (`maritime_adjacency=True`) based on the Marine Regions Maritime Boundaries Geodatabase (Flanders Marine Institute, 2023). Countries are considered adjacent when their exclusive economic zones (EEZs) share a recorded maritime boundary. For joint regimes and overlapping claims, all parties are considered adjacent to one another and to any country whose EEZs border the area. Territories are mapped to their corresponding World Bank country units, while separately listed territories are retained as distinct units. The full maritime layer contains 387 pairs and is combined with land adjacency after inland-water border filtering (Table 2).

**Table 2. Numbers of adjacent pairs under alternative adjacency settings.**

| Spatial level | Lenient (default): all verified borders | Moderate: inland water-only borders require fixed crossings | Stringent: inland water-only borders excluded |
|---|---|---|---|
| Country (ADM0), land only | 326 | 321 | 307 |
| Country (ADM0), land or sea | 583 | 580 | 567 |
| First administrative level (ADM1) | 8,462 | 8,026 | 7,638 |

Note: All counts include de facto adjacency (`de_facto_borders=True`). Land adjacency includes river and lake borders. The land-or-sea row combines the land pairs under each setting with the 387 maritime pairs (`maritime_adjacency=True`).

### *3.3 Inland water boundaries: screens, adjudication, and fixed crossings*

Water-only borders are identified through reproducible spatial screening followed by case-by-case adjudication. Because the hydrographic layers and administrative boundary data lack common identifiers, screening is performed using edge-specific spatial overlays. For each screened pair of ADM1 units (A and B), a border arc is constructed from the corrected boundary polygons. It combines shared boundary segments merged into continuous lines with segments identified where the polygons overlap or where the boundaries face each other across gaps of up to 1,000 m. Hydrographic coverage is defined as the proportion of sample points along the arc that lie within a specified geodesic distance of a mapped river or lake. Sampling uses a nominal maximum interval of 500 m along each continuous arc component.

Seven spatial screens assess potential water-only borders along identifiable ADM1 border arcs. Four screens assess individual hydrographic layers, while three evaluate combinations of river and lake layers (Table 3). The screens identify 3,291 candidate pairs, which undergo a two-stage and two-model evaluation. In the research stage, OpenAI's GPT-5.6 Sol examines each candidate using web sources without access to previous verdicts. In the subsequent adversarial review, Anthropic's Claude Sonnet 5 evaluates the research verdict and spatial coverage estimates, seeking evidence that could challenge the initial classification.

Both passes follow a common classification rubric: a border is classified as water-only when its entire length follows qualifying water features. Any genuine land-defined segment, such as a surveyed line, ridge segment, or overland connector, excludes a water-only classification regardless of its length. Small positional differences between mapped boundaries and waterways may, however, reflect mapping inaccuracies or differences in feature representation rather than genuine land-defined segments. These discrepancies are checked against supporting geographic evidence to avoid rejecting a valid water-only classification. A candidate is accepted when both passes agree on a water-only classification with high confidence and no flagged concerns, or when the database maintainer accepts it following manual review. Candidates rejected by both passes at high confidence are excluded, while disagreements, lower-confidence assessments, and flagged cases are referred to the maintainer for adjudication.

Of the 3,291 candidates, 798 were accepted as water-only borders, seven were identified as false contacts, and 2,486 were rejected. Among the 798 accepted pairs, 579 were nominated by multiple hydrographic layers, 183 exclusively by HydroRIVERS (primarily representing rivers absent from Natural Earth's named river layer), 35 exclusively by the Natural Earth river layer, and one exclusively by HydroLAKES. These results highlight the complementary coverage of the hydrographic datasets and the importance of combining multiple sources to identify water-only borders.

**Table 3. Spatial screens for identifying candidate water-only ADM1 borders.**

| Screen | Features | Base distance | Coverage threshold | Additional distances | Nominated | Accepted |
|---|---|---|---|---|---|---|
| Natural Earth rivers | Named river centerlines | 2.5 km | 0.50 | 5, 10, 15, 20 km | 1,974 | 553 |
| Natural Earth lakes | Lake polygons | 125 m | 0.40 | 250, 500, 1,000, 1,500 m | 170 | 115 |
| HydroRIVERS | All mapped reaches | 500 m | 0.50 | None | 2,048 | 681 |
| HydroLAKES | Lake polygons | 500 m | 0.40 | None | 293 | 156 |
| Combined rivers | Natural Earth rivers or HydroRIVERS | 2.5 km or 500 m | 0.50 | None | 2,269 | 732 |
| Combined lakes | Natural Earth lakes or HydroLAKES | 125 m or 500 m | 0.40 | None | 299 | 159 |
| Cross-type union | All four layers | Maximum per layer (20 km, 1,500 m, or 500 m) | 0.80 | None | 2,050 | 758 |
| Any screen | | | | | 3,291 | 798 |

Note: Screening distances are measured geodesically. The coverage threshold specifies the minimum proportion of border sample points that fall within the specified distance of a mapped water feature for a pair to be nominated. Combined screens count a sample point when it meets the distance criterion for any included layer. Accepted pairs are nominated pairs classified as water-only following adjudication. Counts overlap across screens. 'Any screen' reports unique pairs.

Some water boundaries in the source dataset follow shorelines rather than median lines, leaving the intervening water surface outside both administrative polygons. A separate corridor census therefore identifies potential water-only borders missing from the initial contact graph. The census examines 23,485 pairs within 100 km of each other, excluding existing edges and previously reviewed false contacts while retaining previously recovered water-only borders. Transects connect facing sections of the two outlines, with starting points spaced no more than 250 m apart. A water-proximity score is calculated as the proportion of total transect length within the combined water buffers. A pair is nominated for review if the score is at least 0.80 or either of two narrow-gap criteria is met. The first requires a minimum gap of up to 1 km and at least one transect passing within 500 m of a HydroRIVERS reach. The second requires a minimum gap of up to 5 km and consecutive covered transects (each with at least 80% of its length within the combined water buffers) spanning more than 1 km along either outline. The census nominated 368 pairs, which underwent the same two-model evaluation and maintainer review. Of these, 26 were classified as water-only borders and added as edges in Stage 3, one was identified as a mixed land–water border and added as a land border in Stage 4, and 341 were rejected.

A fixed crossing is defined as a structurally complete road or railway bridge, causeway, dam-top road, or tunnel connecting both administrative units, while ferries, fords, and footbridges are excluded. Fixed-crossing status is assessed for all 824 water-only ADM1 pairs using a four-stage procedure. First, a reproducible screen uses the Overpass API to retrieve mapped bridge features from OpenStreetMap (Haklay and Weber, 2008). A candidate way is retained only if it passes within 100 m of both administrative polygons. Second, a research pass using GPT-5.6 Sol, blinded to the existing classification, determines whether a qualifying crossing exists and reports its name and coordinates. Cases with insufficient evidence receive a negative classification. Third, an adversarial review using Claude Sonnet 5 evaluates the research findings with knowledge of the existing crossing flag. This review examines potential errors, including crossings that connect units other than the target pair, cross water entirely within a single unit, represent excluded crossing types, or remain under construction. Fourth, a reproducible spatial check determines whether the reported crossing location lies within 500 m of the shared border or within 750 m of both units. An existing crossing classification is retained without maintainer review only when both model passes confirm it with high confidence and neither raises a concern, a condition met by 401 pairs. The database maintainer reviewed the remaining 423 pairs and adjudicated disputed or inconsistent classifications. Of the 824 water-only ADM1 pairs, 388 have at least one qualifying fixed crossing.

### *3.4 Validation and interpretation*

Among the candidates identified by the hydrographic screens and corridor census, 263 were automatically accepted and 1,864 automatically rejected based on agreement between both models at high confidence, without flagged concerns. All remaining candidates underwent maintainer review. We assessed the accuracy of the water-only border classifications through human review of samples from predefined groups. The maintainer assessed 80 of the automatically accepted pairs and 80 of the automatically rejected candidates. After excluding four pairs involving World Bank placeholder units, the observed precision among accepted pairs was 98.7% (75/76). Among rejected pairs, the observed false-omission rate was 1.28% (1/78), after excluding two pairs including a placeholder unit and non-touching units, respectively. To assess inter-rater reliability of manually reviewed outcomes, a second rater, blinded to the first maintainer's assessments, independently evaluated 30 sampled pairs. The raters agreed on 28 of 30 pairs (Cohen's $\kappa$ = 0.867; 95% confidence interval: 0.69–1.00).

Coupling-type classification depends on the focal system, spatial scale, and adjacency criteria. A pericoupling relationship at the country level does not imply the same relationship at the ADM1 level, or vice versa. Where relevant, the package provides adjacency validation results at both spatial scales. When the research question or identified systems refer to a city, watershed, protected area, or comparable unsupported unit, no map is generated. The report explains the reason for its omission. These measures reduce the risk of misleading geographic representations.

## 4. LLM-assisted metacoupling workflow

### *4.1 Prompt and parser design*

The qualitative workflow uses a six-layer prompt builder: role, framework-knowledge, methodology, examples, output-format, and interaction-guidelines layers. The framework-knowledge layer defines the coupling types and the five components of the metacoupling framework. The output-format layer establishes a standardized report structure that begins with a coupling classification, followed by separate analyses of intracoupling, pericoupling, and telecoupling, each addressing the five framework components. The report concludes with an assessment of cross-coupling interactions, research gaps, and evidence coverage. The interaction-guidelines layer defines the interaction protocol, instructing the model to generate a complete report in its initial response and refine the existing analysis in subsequent interactions. When classifications are uncertain, the model is instructed to provide a rationale and consider plausible alternatives. Finally, the model response is parsed into a `ParsedAnalysis` data structure, while the original prompt and response remain available for inspection and traceability.

### *4.2 Illustrative application: palm oil production and exports in Indonesia*

The repository includes a recorded execution of the query "Impact of palm oil production and exports in Indonesia on sustainability". The analysis used the OpenAI adapter with gpt-6.1-sol and maximum reasoning effort (`reasoning_effort='max'`). Adjacency lookups used the lenient setting (`coupling_standard='lenient'`), with maritime adjacency (`maritime_adjacency=True`) and de facto adjacency (`de_facto_borders=True`) enabled. Native web search was enabled (`web_search=True`), with a maximum of 25 results (`web_search_max_results=25`). Structured web extraction (`web_structured_extraction=True`) used an additional model call to extract country and flow information for map generation and evidence cards for the qualitative analysis. Literature grounding retrieved up to eight passages from the compiled corpus (`rag_top_k=8`), with a maximum of five passages per publication (`rag_max_chunks_per_paper=5`). Automatic map generation was enabled (`auto_map=True`). The execution comprised five model calls: native web search, structured web extraction, framework analysis, map-data extraction, and abstract generation. The built-in execution trace captured the final three calls, totaling 79,305 tokens, and outputs from the two web-stage calls were retained separately.

In response to the free-text query, the model identified Indonesia as the national focal system and organized the analysis into three coupling types (Fig. 2). Intracoupling encompassed plantation management, processing, domestic food and biodiesel consumption, and associated environmental and livelihood effects. Pericoupling included exports to India, the associated embodied water, the corresponding payments, and trade-related information. Telecoupling linked Indonesia with China, the European Union, the United States, Pakistan, and Switzerland through product exports, payments, and policy-related information. The report connected production to forest conversion, peat and water pressures, employment, and uneven livelihood outcomes. It also identified interactions between domestic biodiesel demand and export markets. Malaysia, the other major producer, appeared as a candidate spillover system: shifts in buyers' sourcing from Indonesia could affect Malaysian producers.

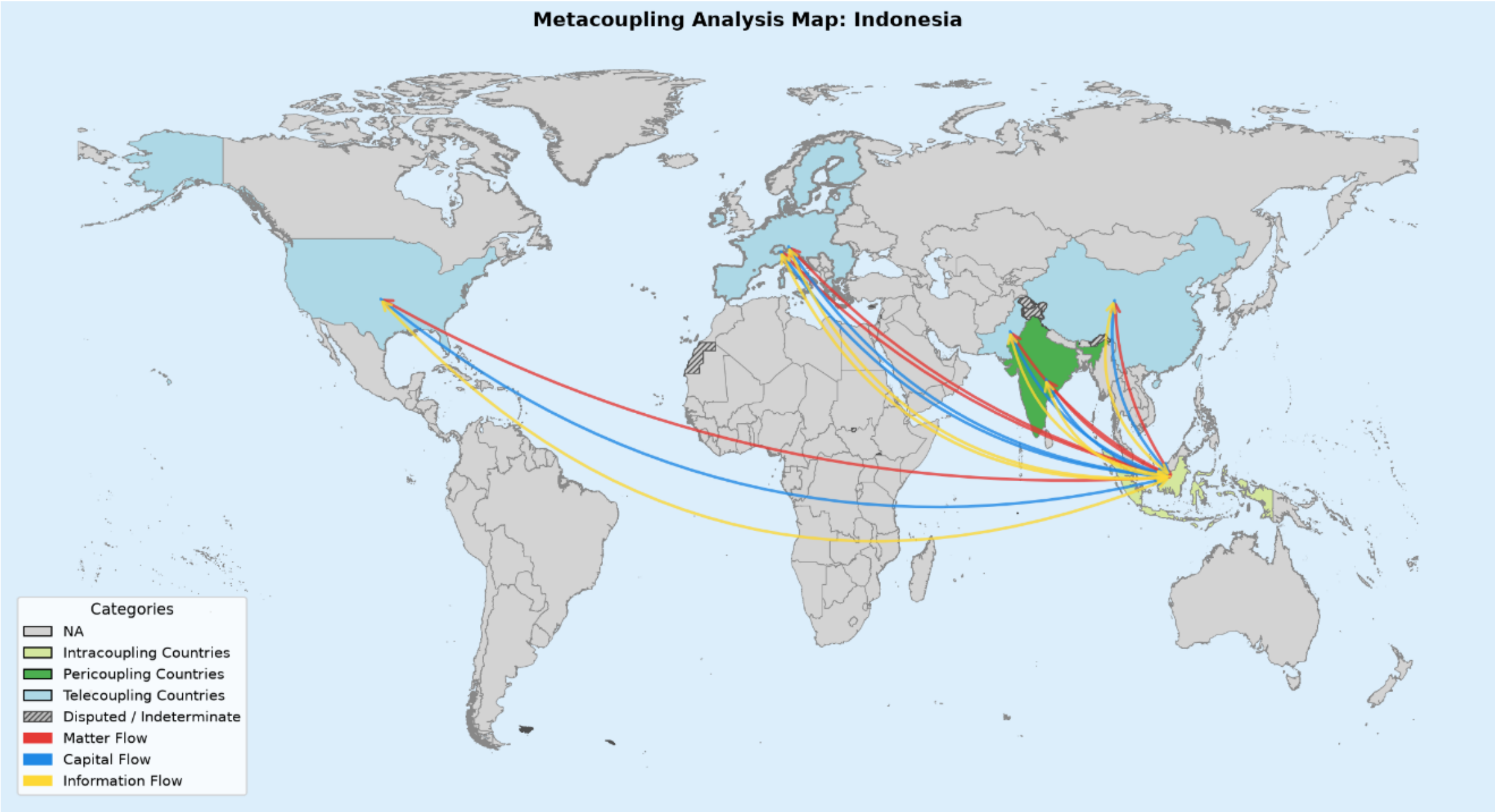

**Fig. 2. Illustrative metacoupling map of palm oil production and exports in Indonesia.** Rendered by MetacouplingLLM. Country shading indicates coupling types relative to Indonesia. Arrows show palm-product exports, reverse capital flows, and information flows. India is classified as pericoupled through maritime adjacency. The European Union is shown as a collective market, and its arrows meet at a single anchor point, next to those for Switzerland.

### *4.3 Adjacency validation, citations, and evidence coverage*

Two mechanisms support the validation of model-generated outputs. First, adjacency validation operates after parsing and does not require additional model calls. It matches the countries named as systems in the analysis to the database and uses the adjacency database to determine whether each pair involving the focal system is pericoupled or telecoupled, according to the user-specified `de_facto_borders`, `coupling_standard`, and `maritime_adjacency` settings. The report warns users when an adjacent partner is identified but pericoupling is omitted, when all checked partners are non-adjacent but telecoupling is omitted, or when a partner appears only in the coupling section that its database classification contradicts. Partners named in both the pericoupling and telecoupling sections are not assessed for placement, and spillover systems are not validated.

The second mechanism provides citation and evidence tracking. Literature passages are cited as [Tk:N] and web sources as [Tk:Wn], where k is the turn (for example, [T1:3] and [T1:W5] in the first turn). A citation sanitizer removes markers that cannot be matched to retrieved sources, and the report concludes with reference lists linking each retained marker to its corresponding publication or web page. Each report also includes an evidence coverage note that summarizes the model's assessment of how well the retrieved evidence supports different groups of claims.

Country-level validation classified Indonesia–India as pericoupled and Indonesia's links with China, the European Union, the United States, Pakistan, and Switzerland as telecoupled, consistent with the model's analysis. The Indonesia–India relationship demonstrates the maritime layer's contribution: the countries share a maritime boundary but no land border. With maritime adjacency disabled (`maritime_adjacency=False`), the same pair is classified as telecoupled; enabling the layer changes its geographic classification without changing the trade evidence. The run retrieved eight literature passages and 25 web sources and the model cited six of the passages and 20 of the web sources. The model-generated evidence coverage note identified the strongest support for plantation expansion, forest conversion, and environmental pressures and for heterogeneous livelihood outcomes. It characterized the trade evidence as direct but uneven across partners, while distinguishing it from inferred or unverified claims about payment distributions, information networks, land-use effects in importing countries, and Malaysia's proposed spillover response.

## 5. Metacoupled flow indicators

We propose three quantitative indicators for characterizing metacoupled flows: Metacoupled Flow Shares (MFS), Metacoupled Flow Evenness (MFE), and the Metacoupled Flow Concentration Index (MFCI). The indicators are calculated from user-supplied flow magnitudes and coupling-type labels, with the latter optionally derived from the adjacency databases. For a given set of inputs, the resulting indicator values are reproducible.

### *5.1 Quantitative indicators*

The package operationalizes the metacoupling framework (Liu, 2017) using a flow table in which each row records a directed flow from focal system $i$ to partner $j$, which is attributed to intracoupling ($I$, within the focal system), pericoupling ($P$, to an adjacent system), or telecoupling ($T$, to a distant system).

Metacoupled Flow Shares (MFS) quantify the relative contributions of intracoupled, pericoupled, and telecoupled flows to a focal system's total flow. For each coupling type $c \in \{I, P, T\}$, the flow share $\mathrm{MFS}_i^c$ is the proportion of the total flow of focal system $i$ attributed to that coupling type; the three shares range from 0 to 1 and sum to 1:

$$\begin{aligned} \mathrm{MFS}_i^c &= \frac{F_i^c}{F_i}, \\ F_i^c &= \sum_j f_{ij}^c, \\ F_i &= F_i^I + F_i^P + F_i^T \end{aligned} \qquad (1)$$

where $f_{ij}^c$ is the type-$c$ flow from focal system $i$ to partner $j$, $F_i^c$ is the total type-$c$ flow of system $i$, and $F_i$ is its total flow. The intracoupled, pericoupled, and telecoupled flow shares equal to $\mathrm{MFS}_i^I$, $\mathrm{MFS}_i^P$, and $\mathrm{MFS}_i^T$, respectively.

Metacoupled Flow Evenness (MFE) measures how evenly flow is distributed across intracoupling, pericoupling, and telecoupling, drawing on Shannon's information-entropy formulation (Shannon, 1948) and Pielou's entropy-based approach to evenness (Pielou, 1966). The entropy of their flow shares is normalized by its maximum under equal flow shares. MFE ranges from 0, when all flow is confined to one coupling type, to 1, when each coupling type accounts for one-third of total flow. By convention, when a share equals 0, $0 \ln 0 = 0$.

$$\mathrm{MFE}_i = -\frac{1}{\ln 3} \sum_{c \in \{I, P, T\}} \mathrm{MFS}_i^c \ln \mathrm{MFS}_i^c \qquad (2)$$

The Metacoupled Flow Concentration Index (MFCI) quantifies how unevenly a focal system's flow is distributed among partners within each coupling type, yielding intracoupled, pericoupled, and telecoupled concentration indices (IFCI, PFCI, and TFCI). It is a framework-specific adaptation of the Herfindahl–Hirschman concentration index (Hirschman, 1945; Herfindahl, 1950), with normalization relative to equal flow shares among the observed partners. For coupling types containing at least two partners, values range from 0 under equal flow shares toward 1 as one partner increasingly dominates. Coupling types with a single partner are assigned a value of 1 by convention.

$$
\begin{aligned}
\mathrm{HHI}_i^c &= \sum_j \left(\frac{f_{ij}^c}{F_i^c}\right)^2, \\
\mathrm{MFCI}_i^c &= \frac{\mathrm{HHI}_i^c - 1/n_i^c}{1 - 1/n_i^c}
\end{aligned}
\tag{3}
$$

where $n_i^c$ is the number of partners with type-$c$ flow.

The package also reports the Equivalent Number of Partners (ENP), which translates the concentration into an intuitive count (Laakso and Taagepera, 1979). For example, a system whose telecoupled flow is split among five partners of unequal size might have $\mathrm{ENP}^T \approx 3.2$, meaning its concentration is equivalent to that of 3.2 equally sized partners.

$$
\mathrm{ENP}_i^c = \frac{1}{\mathrm{HHI}_i^c} \tag{4}
$$

$\mathrm{ENP}_i^c$ is reported per coupling type as $\mathrm{ENP}^I$, $\mathrm{ENP}^P$, and $\mathrm{ENP}^T$, the equivalent numbers of intracoupled, pericoupled, and telecoupled partners.

When total flow across all coupling types is zero, MFS, MFE, MFCI, and ENP are all undefined. When there is no flow for a specific coupling type, the coupling-specific MFCI and ENP are undefined.

### *5.2 Illustrative application: global marine fishing across exclusive economic zones*

Carlson et al. (2020) quantified fish catches within EEZs from 1950 to 2014 and classified them according to intracoupling, pericoupling, and telecoupling relationships. Their analysis examined the spatial and temporal patterns of these coupling types at global, regional, and individual-EEZ scales, revealing their relative contributions and changes over time. In this application, we extend their analysis with MFS, MFE, and MFCI, which characterize the distribution of catches among coupling types and among fishing nations.

Following Carlson et al. (2020), we treated each of the 280 EEZs as a focal system and assigned catches to intracoupling, pericoupling, or telecoupling. We used the national database under the lenient setting (`coupling_standard='lenient'`) with maritime adjacency enabled (`maritime_adjacency=True`), matching EEZs and fishing nations to its country units (Section 3.2). Adjacency was evaluated between these country units, and the same geographic configuration was applied to pooled 1950–2014 catches. Table 4 summarizes regional catch totals and proportions for 1950–2014 based on data from Carlson et al. (2020), with Types 1, 2, and 3 corresponding to intracoupling, pericoupling, and telecoupling, respectively. Telecoupling (Type 3) accounted for the largest share of cumulative catches in West Africa (54.3%) and Oceania (53.2%). Intracoupling (Type 1) dominated in the remaining nine regions, with the highest shares observed in Central America (97.0%), South America (94.2%), and the Caribbean (91.6%).

**Table 4. Regional marine catch totals and shares by coupling type, 1950–2014.**

| Region | n | Total (MMT) | Type 1 (MMT) | Type 2 (MMT) | Type 3 (MMT) | Type 1 (%) | Type 2 (%) | Type 3 (%) |
|---|---|---|---|---|---|---|---|---|
| Africa (West) | 13 | 286.4 | 118.7 | 12.1 | 155.6 | 41.4 | 4.2 | 54.3 |
| Oceania | 26 | 100.5 | 44.0 | 3.0 | 53.4 | 43.8 | 3.0 | 53.2 |
| Africa (non-western) | 30 | 433.4 | 209.8 | 55.6 | 168.0 | 48.4 | 12.8 | 38.8 |
| Europe (E/S) | 31 | 156.5 | 97.1 | 20.7 | 38.8 | 62.0 | 13.2 | 24.8 |
| North America | 19 | 575.1 | 421.6 | 27.1 | 126.4 | 73.3 | 4.7 | 22.0 |
| Europe (North) | 25 | 545.2 | 287.7 | 181.6 | 75.9 | 52.8 | 33.3 | 13.9 |
| Europe (West) | 29 | 126.0 | 105.4 | 15.4 | 5.3 | 83.6 | 12.2 | 4.2 |
| Caribbean | 14 | 13.9 | 12.8 | 0.8 | 0.4 | 91.6 | 5.5 | 2.9 |
| South America | 18 | 840.4 | 791.9 | 32.0 | 16.5 | 94.2 | 3.8 | 2.0 |
| Central America | 14 | 123.0 | 119.3 | 2.4 | 1.3 | 97.0 | 2.0 | 1.0 |
| Asia | 61 | 2532.5 | 2099.4 | 410.3 | 22.8 | 82.9 | 16.2 | 0.9 |

Note: Calculated from Carlson et al. (2020) catch data using MetacouplingLLM adjacency classifications. MMT denotes million metric tons. Values are cumulative catches for 1950–2014. n is the number of distinct EEZs. Types 1, 2, and 3 denote intracoupling, pericoupling, and telecoupling, respectively.

Using these reclassified flows, we calculated the metacoupled flow indicators across 11 regions (Table 5). Indicators are total-catch-weighted means of EEZ-level values from pooled 1950–2014 catches. IFCI and $ENP^{I}$ are omitted because domestic catches are aggregated to a single national partner, giving both a value of 1 wherever domestic catch is positive.

**Table 5. Catch-weighted regional means of metacoupled flow indicators, 1950–2014.**

| Region | n | IFS | PFS | TFS | MFE | PFCI | TFCI | $ENP^{P}$ | $ENP^{T}$ |
|---|---|---|---|---|---|---|---|---|---|
| Africa (West) | 13 | 0.41 | 0.04 | 0.54 | 0.54 | 0.63 | 0.20 | 1.50 | 8.20 |
| Oceania | 26 | 0.44 | 0.03 | 0.53 | 0.55 | 0.76 | 0.27 | 1.34 | 3.54 |
| Africa (non-western) | 30 | 0.48 | 0.13 | 0.39 | 0.65 | 0.76 | 0.24 | 1.21 | 5.28 |
| Europe (E/S) | 31 | 0.62 | 0.13 | 0.25 | 0.47 | 0.47 | 0.40 | 2.20 | 2.90 |
| North America | 19 | 0.73 | 0.05 | 0.22 | 0.49 | 0.69 | 0.42 | 1.43 | 3.64 |
| Europe (North) | 25 | 0.53 | 0.33 | 0.14 | 0.69 | 0.36 | 0.27 | 2.65 | 3.38 |
| Europe (West) | 29 | 0.84 | 0.12 | 0.04 | 0.30 | 0.53 | 0.32 | 1.79 | 3.19 |
| Caribbean | 14 | 0.92 | 0.06 | 0.03 | 0.24 | 0.79 | 0.39 | 1.22 | 2.46 |
| South America | 18 | 0.94 | 0.04 | 0.02 | 0.21 | 0.94 | 0.26 | 1.04 | 3.11 |
| Central America | 14 | 0.97 | 0.02 | 0.01 | 0.12 | 0.97 | 0.52 | 1.02 | 1.88 |
| Asia | 61 | 0.83 | 0.16 | 0.01 | 0.36 | 0.61 | 0.43 | 1.67 | 2.52 |

Note: Mean MFCI and ENP values were calculated using only EEZs with positive catches in the corresponding coupling type, with weights renormalized across eligible EEZs.

The IFS, PFS, and TFS values in Table 5 represent the same regional coupling proportions as Table 4, expressed as fractions, whereas MFE, PFCI, TFCI, and ENP are catch-weighted means of EEZ-level values. Northern European EEZs have the highest mean MFE (0.69), indicating that their catches are the most evenly distributed among coupling types, whereas the regions with the lowest mean MFE (Central America, South America, and the Caribbean) are dominated by a single coupling type. PFCI, TFCI, and ENP further characterize the concentration of catches among fishing nations within each coupling type. For example, West Africa and Oceania have similar telecoupled flow shares (0.54 and 0.53, respectively), but Oceania has a higher TFCI (0.27) than West Africa (0.20), indicating greater concentration of telecoupled catches. This difference is also reflected in $ENP^T$, which is equivalent to 3.54 equally sized distant fishing nations for Oceanian EEZs, compared with 8.20 for West African EEZs. Across the 269 EEZs with positive distant catches, the unweighted mean equivalent number of distant fishing nations was 3.15, compared with a mean nominal count of 17.60. This difference indicates that telecoupled catches were unevenly distributed among distant fishing nations.

### *5.3 LLM assistance around the reproducible indicators*

MetacouplingLLM provides five optional LLM helpers for study definition, input assessment, ambiguous-flow classification, result interpretation, and methods drafting. Indicator values are calculated reproducibly from the supplied flow magnitudes and coupling-type labels. Each helper returns an `LLMTrace` object recording the model identifier, prompts, response, and timestamp. The classification helper additionally provides model-assigned confidence assessments, classification rationales, and confirmation flags. Given fixed input data and coupling-type labels, repeated calculations yield the same indicator values, whereas LLM-generated classifications and text may vary across runs.

## 6. Discussion

MetacouplingLLM connects qualitative research framing with geographic classification and quantitative flow analysis. This integration makes the relationship between study-specific assumptions and analytical results more explicit. The LLM-assisted workflow provides a structured starting point that researchers can review and refine, while the adjacency databases and indicator functions support analyses of user-supplied flow data using consistent definitions. Researchers can also examine how alternative adjacency settings change coupling labels and indicator values while holding the flow data fixed. These capabilities complement the Telecoupling Toolbox and GeoApp, which support spatial analyses of user-defined systems using supplied geographic data (Tonini and Liu, 2017; McCord et al., 2018). MetacouplingLLM extends this support to preliminary research framing and provides reusable methods for comparing flow distributions and partner concentration.

The pericoupling database addresses a gap between existing geographic datasets and the information required to classify interactions between adjacent systems. Although geographic adjacency is fundamental to pericoupling classification, no single dataset adequately captures all relevant boundaries. The World Bank Official Boundaries dataset provides country and first-level administrative (ADM1) geometries but lacks inland-water features. Some river boundaries are represented along opposite banks, leaving the intervening water surface outside both administrative units. In addition, small polygon slivers may create false contacts or obscure actual shared boundaries. Natural Earth provides river and lake features, but its river centerlines at 1:10,000,000 scale may deviate by several kilometers from administrative boundaries following the same rivers, and many smaller water bodies are omitted. HydroRIVERS, derived from terrain data at 15-arc-second resolution, provides additional stream coverage, although its modeled river reaches may not align precisely with administrative boundaries. HydroLAKES further supplements the coverage of smaller lakes. The maritime layer uses Marine Regions data to add country-level adjacency based on shared EEZ boundaries, which enables researchers to include adjacency at sea when relevant to their study. The database integrates these complementary sources to address their individual limitations rather than relying on any single dataset for classification. Together, these resources provide a configurable and reliable geographic reference.

Fixed-crossing identification adds an infrastructure dimension to geographic adjacency by distinguishing water-only borders with qualifying road or rail crossings from those without identified crossings. Boundary and hydrographic datasets alone cannot establish the presence of these connections, making crossing verification an important complement to spatial screening. The resulting records provide an evidence base for the optional moderate adjacency standard and enable researchers to assess how crossing-dependent classifications affect pericoupled and telecoupled flow indicators. OpenStreetMap provides globally available and openly licensed information on bridges, tunnels, and transportation infrastructure, but the completeness and accuracy of these volunteered data vary geographically (Haklay, 2010; Barrington-Leigh and Millard-Ball, 2017). Mapped features may represent nonqualifying footbridges or unfinished crossings, while existing crossings may be absent from the data. The package therefore integrates OpenStreetMap screening, two-model adjudication, and reproducible spatial verification, with unresolved cases reviewed by the maintainer and supporting evidence documented for each decision. This approach reduces reliance on incomplete or ambiguous infrastructure records when determining whether water-only borders qualify under the moderate setting.

This study has three principal limitations: the evaluation of qualitative outputs, reliance on external models, and the scope and temporal currency of the geographic database. First, the accuracy and completeness of model-generated qualitative analyses have not been systematically evaluated. The package provides several safeguards: adjacency validation checks coupling types against the database, citation sanitization removes references to sources that were not retrieved, and the evidence coverage note provides the model's self-assessment of supporting evidence. However, these mechanisms do not establish whether the identified systems, flows, agents, causes, and effects are accurate and complete, or whether the cited sources adequately support individual claims. Future work will develop an evaluation dataset covering different coupling types, research contexts, and spatial scales. Independent experts will assess generated analyses using a predefined rubric covering factual accuracy, framework completeness, and evidence support, informed by expert-prepared reference analyses. Evaluation scores and inter-rater agreement will be reported.

Second, the LLM-assisted workflow relies on commercial models whose behavior and availability are beyond the package's control. Providers frequently introduce, update, and retire models (Ollion et al., 2024), and model performance may change over time even when the model identifier remains the same (Ma et al., 2024). Future work will benchmark locally deployable and open-weight models against commercial models through expert assessments of factual accuracy, framework consistency, and evidence support. Suitable local models will be integrated as alternatives, with fine-tuning on metacoupling-specific data considered to address weaknesses identified during evaluation.

Finally, the geographic database represents adjacency and fixed-crossing presence but does not provide a comprehensive assessment of functional connectivity. Shared land borders do not account for terrain-related barriers or accessibility through transport networks. Similarly, fixed-crossing classifications do not capture crossing capacity, utilization, or connections to population and production centers. Their relevance also varies by flow type, because infrastructure supporting human mobility and trade does not necessarily facilitate ecological or hydrological flows. Current adjacency databases can provide a foundation for assessing interaction intensity, but they alone cannot fully capture it. The database also reflects conditions at the time of compilation. Current fixed-crossing classifications draw on OpenStreetMap screening and web evidence collected in September and October 2026 and may become outdated as crossings are constructed, removed, or modified. Administrative boundaries and their source datasets also change over time. Future work could supplement adjacency classifications with flow-specific measures of accessibility, while retaining geographic adjacency as a separate classification criterion. Regular maintenance should include periodic reassessment, targeted review of reported changes, and a public correction process, with evidence sources and verification dates retained alongside previous database versions.

## 7. Conclusion

MetacouplingLLM provides an open-source computational framework for applying the metacoupling framework through geographic classification, quantitative flow indicators, and LLM-assisted qualitative analysis. By distinguishing reproducible calculations from model-generated interpretations, the package supports both researchers working with existing flow datasets and those developing new research designs. This study contributes an integrated adjacency database and framework-specific flow indicators, combined with an LLM-assisted workflow. Together, these capabilities make analytical assumptions more explicit and support systematic characterization of intracoupling, pericoupling, and telecoupling.

The pericoupling databases integrate administrative boundaries, hydrographic evidence, and reviewed corrections into a documented geographic reference. Fixed-crossing identification further distinguishes water-only borders based on qualifying infrastructure, while alternative adjacency standards allow researchers to examine how different classification assumptions affect their results. MFS, MFE, and MFCI complement this geographic foundation by quantifying flow composition, evenness across coupling types, and concentration among partners, respectively. The marine fishing application shows that regions with similar telecoupled flow shares may exhibit substantially different levels of catch concentration among fishing nations, highlighting information not captured by flow shares. The palm oil application illustrates how literature retrieval, web evidence, and geographic validation support a structured metacoupling analysis.

These applications demonstrate the package's practical functionality, although further evaluation is needed to assess the accuracy and consistency of its qualitative output. Future priorities include systematic expert assessment of LLM-generated analyses, comparative evaluation of locally deployable LLMs, and flow-specific accessibility measures, alongside regular updates to adjacency databases. MetacouplingLLM is intended to complement empirical investigation and expert judgment. By providing reusable analytical tools and explicit geographic assumptions, the package supports transparent and comparative research on interactions within and across coupled human and natural systems.

## References

Anselin, L., Syabri, I., Kho, Y., 2006. GeoDa: An introduction to spatial data analysis. Geogr. Anal. 38, 5–22. https://doi.org/10.1111/j.0016-7363.2005.00671.x.

Babaei Giglou, H., D'Souza, J., Auer, S., 2024. LLMs4Synthesis: Leveraging large language models for scientific synthesis, in: Proceedings of the 24th ACM/IEEE Joint Conference on Digital Libraries. ACM, New York, Article 31, pp. 1–12. https://doi.org/10.1145/3677389.3702565.

Barrington-Leigh, C., Millard-Ball, A., 2017. The world's user-generated road map is more than 80% complete. PLoS One 12, e0180698. https://doi.org/10.1371/journal.pone.0180698.

Bicudo da Silva, R.F., Viña, A., Moran, E.F., Dou, Y., Batistella, M., Liu, J., 2021. Socioeconomic and environmental effects of soybean production in metacoupled systems. Sci. Rep. 11, 18662. https://doi.org/10.1038/s41598-021-98256-6.

Carlson, A.K., Taylor, W.W., Rubenstein, D.I., Levin, S.A., Liu, J., 2020. Global marine fishing across space and time. Sustainability 12, 4714. https://doi.org/10.3390/su12114714.

Chung, M.G., Dietz, T., Liu, J., 2018. Global relationships between biodiversity and nature-based tourism in protected areas. Ecosyst. Serv. 34, 11–23. https://doi.org/10.1016/j.ecoser.2018.09.004.

Dagdelen, J., Dunn, A., Lee, S., Walker, N., Rosen, A.S., Ceder, G., Persson, K.A., Jain, A., 2024. Structured information extraction from scientific text with large language models. Nat. Commun. 15, 1418. https://doi.org/10.1038/s41467-024-45563-x.

Dou, Y., Yao, G., Herzberger, A., Bicudo da Silva, R.F., Song, Q., Hovis, C., Batistella, M., Moran, E., Wu, W., Liu, J., 2020a. Land-use changes in distant places: Implementation of a telecoupled agent-based model. J. Artif. Soc. Soc. Simul. 23, 11. https://doi.org/10.18564/jasss.4211.

Dou, Y., Bicudo da Silva, R.F., McCord, P., Zaehringer, J.G., Yang, H., Furumo, P.R., Zhang, J., Pizarro, J.C., Liu, J., 2020b. Understanding how smallholders integrated into pericoupled and telecoupled systems. Sustainability 12, 1596. https://doi.org/10.3390/su12041596.

Drakou, E.G., Virdin, J., Pendleton, L., 2018. Mapping the global distribution of locally-generated marine ecosystem services: The case of the West and Central Pacific Ocean tuna fisheries. Ecosyst. Serv. 31, 278–288. https://doi.org/10.1016/j.ecoser.2018.05.008.

Du, Y., Zhao, D., Qiu, S., Zhou, F., Peng, J., 2022. How can virtual water trade reshape water stress pattern? A global evaluation based on the metacoupling perspective. Ecol. Indic. 145, 109712. https://doi.org/10.1016/j.ecolind.2022.109712.

Flanders Marine Institute, 2023. Maritime Boundaries Geodatabase: Maritime boundaries and exclusive economic zones (200NM) [dataset]. Version 12. Marine Regions. https://doi.org/10.14284/632.

Gao, Y., Xiong, Y., Gao, X., Jia, K., Pan, J., Bi, Y., Dai, Y., Sun, J., Wang, M., Wang, H., 2023. Retrieval-augmented generation for large language models: A survey. arXiv:2312.10997. https://doi.org/10.48550/arXiv.2312.10997.

Geng, J., Cai, F., Wang, Y., Koeppl, H., Nakov, P., Gurevych, I., 2024. A survey of confidence estimation and calibration in large language models, in: Proceedings of the 2024 Conference of the North American Chapter of the Association for Computational Linguistics: Human Language Technologies (Volume 1: Long Papers). Association for Computational Linguistics, Mexico City, pp. 6577–6595. https://doi.org/10.18653/v1/2024.naacl-long.366.

Haklay, M., 2010. How good is volunteered geographical information? A comparative study of OpenStreetMap and Ordnance Survey datasets. Environ. Plann. B Plann. Des. 37, 682–703. https://doi.org/10.1068/b35097.

Haklay, M., Weber, P., 2008. OpenStreetMap: User-generated street maps. IEEE Pervasive Comput. 7, 12–18. https://doi.org/10.1109/MPRV.2008.80.

Herfindahl, O.C., 1950. Concentration in the Steel Industry. PhD thesis, Columbia University, New York.

Herzberger, A., Chung, M.G., Kapsar, K., Frank, K.A., Liu, J., 2019. Telecoupled food trade affects pericoupled trade and intracoupled production. Sustainability 11, 2908. https://doi.org/10.3390/su11102908.

Hirschman, A.O., 1945. National Power and the Structure of Foreign Trade. University of California Press, Berkeley.

Hovis, C.L., Dou, Y., Herzberger, A., Liu, J., 2021. Through the lens of telecoupling and metacoupling: New perspectives for global sustainability. Sustainability 13, 2953. https://doi.org/10.3390/su13052953.

Hull, V., Liu, J., 2018. Telecoupling: A new frontier for global sustainability. Ecol. Soc. 23, 41. https://doi.org/10.5751/ES-10494-230441.

Ji, Z., Lee, N., Frieske, R., Yu, T., Su, D., Xu, Y., Ishii, E., Bang, Y.J., Madotto, A., Fung, P., 2023. Survey of hallucination in natural language generation. ACM Comput. Surv. 55, 248. https://doi.org/10.1145/3571730.

Kalai, A.T., Vempala, S.S., 2024. Calibrated language models must hallucinate, in: Proceedings of the 56th Annual ACM Symposium on Theory of Computing. ACM, New York, pp. 160–171. https://doi.org/10.1145/3618260.3649777.

Kapsar, K.E., Hovis, C.L., Bicudo da Silva, R.F., Buchholtz, E.K., Carlson, A.K., Dou, Y., Du, Y., Furumo, P.R., Li, Y., Torres, A., Yang, D., Wan, H.Y., Zaehringer, J.G., Liu, J., 2019. Telecoupling research: The first five years. Sustainability 11, 1033. https://doi.org/10.3390/su11041033.

Laakso, M., Taagepera, R., 1979. "Effective" number of parties: A measure with application to West Europe. Comp. Polit. Stud. 12, 3–27. https://doi.org/10.1177/001041407901200101.

Lehner, B., Grill, G., 2013. Global river hydrography and network routing: Baseline data and new approaches to study the world's large river systems. Hydrol. Process. 27, 2171–2186. https://doi.org/10.1002/hyp.9740.

Lewis, P., Perez, E., Piktus, A., Petroni, F., Karpukhin, V., Goyal, N., Küttler, H., Lewis, M., Yih, W., Rocktäschel, T., Riedel, S., Kiela, D., 2020. Retrieval-augmented generation for knowledge-intensive NLP tasks, in: Advances in Neural Information Processing Systems 33. Curran Associates, Red Hook, NY, pp. 9459–9474. https://proceedings.neurips.cc/paper/2020/hash/6b493230205f780e1bc26945df7481e5-Abstract.html.

Liu, J., 2017. Integration across a metacoupled world. Ecol. Soc. 22, 29. https://doi.org/10.5751/ES-09830-220429.

Liu, J., 2023. Leveraging the metacoupling framework for sustainability science and global sustainable development. Natl. Sci. Rev. 10, nwad090. https://doi.org/10.1093/nsr/nwad090.

Liu, J., Dietz, T., Carpenter, S.R., Alberti, M., Folke, C., Moran, E., Pell, A.N., Deadman, P., Kratz, T., Lubchenco, J., Ostrom, E., Ouyang, Z., Provencher, W., Redman, C.L., Schneider, S.H., Taylor, W.W., 2007. Complexity of coupled human and natural systems. Science 317, 1513–1516. https://doi.org/10.1126/science.1144004.

Liu, J., Hull, V., Batistella, M., DeFries, R., Dietz, T., Fu, F., Hertel, T.W., Izaurralde, R.C., Lambin, E.F., Li, S., Martinelli, L.A., McConnell, W.J., Moran, E.F., Naylor, R., Ouyang, Z., Polenske, K.R., Reenberg, A., de Miranda Rocha, G., Simmons, C.S., Verburg, P.H., Vitousek, P.M., Zhang, F., Zhu, C., 2013. Framing sustainability in a telecoupled world. Ecol. Soc. 18, 26. https://doi.org/10.5751/ES-05873-180226.

Liu, J., Mooney, H., Hull, V., Davis, S.J., Gaskell, J., Hertel, T., Lubchenco, J., Seto, K.C., Gleick, P., Kremen, C., Li, S., 2015. Systems integration for global sustainability. Science 347, 1258832. https://doi.org/10.1126/science.1258832.

Liu, J., Dou, Y., Batistella, M., Challies, E., Connor, T., Friis, C., Millington, J.D.A., Parish, E., Romulo, C.L., Bicudo da Silva, R.F., Triezenberg, H., Yang, H., Zhao, Z., Zimmerer, K.S., Huettmann, F., Treglia, M.L., Basher, Z., Chung, M.G., Herzberger, A., Lenschow, A., Mechiche-Alami, A., Newig, J., Roche, J., Sun, J., 2018. Spillover systems in a telecoupled Anthropocene: Typology, methods, and governance for global sustainability. Curr. Opin. Environ. Sustain. 33, 58–69. https://doi.org/10.1016/j.cosust.2018.04.009.

Ma, W., Yang, C., Kästner, C., 2024. (Why) is my prompt getting worse? Rethinking regression testing for evolving LLM APIs, in: Proceedings of the IEEE/ACM 3rd International Conference on AI Engineering - Software Engineering for AI (CAIN 2024). ACM, New York, pp. 166–171. https://doi.org/10.1145/3644815.3644950.

Manning, N., Li, Y., Liu, J., 2023. Broader applicability of the metacoupling framework than Tobler's first law of geography for global sustainability: A systematic review. Geogr. Sustain. 4, 6–18. https://doi.org/10.1016/j.geosus.2022.11.003.

McCord, P., Tonini, F., Liu, J., 2018. The Telecoupling GeoApp: A Web-GIS application to systematically analyze telecouplings and sustainable development. Appl. Geogr. 96, 16–28. https://doi.org/10.1016/j.apgeog.2018.05.001.

Messager, M.L., Lehner, B., Grill, G., Nedeva, I., Schmitt, O., 2016. Estimating the volume and age of water stored in global lakes using a geo-statistical approach. Nat. Commun. 7, 13603. https://doi.org/10.1038/ncomms13603.

Millington, J.D.A., Xiong, H., Peterson, S., Woods, J., 2017. Integrating modelling approaches for understanding telecoupling: Global food trade and local land use. Land 6, 56. https://doi.org/10.3390/land6030056.

Natural Earth, 2022. Natural Earth: Free vector and raster map data at 1:10m, 1:50m, and 1:110m scales [dataset]. Version 5.1.2. https://github.com/nvkelso/natural-earth-vector/releases/tag/v5.1.2.

Ollion, É., Shen, R., Macanovic, A., Chatelain, A., 2024. The dangers of using proprietary LLMs for research. Nat. Mach. Intell. 6, 4–5. https://doi.org/10.1038/s42256-023-00783-6.

Pielou, E.C., 1966. The measurement of diversity in different types of biological collections. J. Theor. Biol. 13, 131–144. https://doi.org/10.1016/0022-5193(66)90013-0.

Prell, C., Sun, L., Feng, K., He, J., Hubacek, K., 2017. Uncovering the spatially distant feedback loops of global trade: A network and input-output approach. Sci. Total Environ. 586, 401–408. https://doi.org/10.1016/j.scitotenv.2016.11.202.

Rey, S.J., Anselin, L., 2007. PySAL: A Python library of spatial analytical methods. Rev. Reg. Stud. 37, 5–27. https://doi.org/10.52324/001c.8285.

Schaffer-Smith, D., Tomscha, S.A., Jarvis, K.J., Maguire, D.Y., Treglia, M.L., Liu, J., 2018. Network analysis as a tool for quantifying the dynamics of metacoupled systems: An example using global soybean trade. Ecol. Soc. 23, 3. https://doi.org/10.5751/ES-10460-230403.

Sclar, M., Choi, Y., Tsvetkov, Y., Suhr, A., 2024. Quantifying language models' sensitivity to spurious features in prompt design or: How I learned to start worrying about prompt formatting, in: The Twelfth International Conference on Learning Representations (ICLR 2024). https://openreview.net/forum?id=RIu5lyNXjT.

Shannon, C.E., 1948. A mathematical theory of communication. Bell Syst. Tech. J. 27, 379–423. https://doi.org/10.1002/j.1538-7305.1948.tb01338.x.

Siddiq, M.L., Islam-Gomes, A., Sekerak, N., Santos, J.C.S., 2025. Large language models for software engineering: A reproducibility crisis. arXiv:2512.00651. https://doi.org/10.48550/arXiv.2512.00651.

Tobler, W.R., 1970. A computer movie simulating urban growth in the Detroit region. Econ. Geogr. 46, 234–240. https://doi.org/10.2307/143141.

Tonini, F., Liu, J., 2017. Telecoupling Toolbox: Spatially explicit tools for studying telecoupled human and natural systems. Ecol. Soc. 22, 11. https://doi.org/10.5751/ES-09696-220411.

World Bank, 2026. World Bank Official Boundaries [dataset]. World Bank Development Data Hub, release of 14 May 2026. https://datacatalog.worldbank.org/search/dataset/0038272/world-bank-official-boundaries (accessed 3 June 2026).

Xu, L., Zhao, S., Lin, Q., Chen, L., Luo, Q., Wu, S., Ye, X., Feng, H., Du, Z., 2025. Evaluating large language models on geospatial tasks: A multiple geospatial task benchmarking study. Int. J. Digit. Earth 18, 2480268. https://doi.org/10.1080/17538947.2025.2480268.